\documentclass[runningheads]{llncs}
\usepackage[T1]{fontenc}
\usepackage{tabularx}
\usepackage{graphicx}
\usepackage{amsmath}
\usepackage{booktabs}
\usepackage{amsfonts}
\usepackage{multirow}
\usepackage{pgfplots} 
\usetikzlibrary{patterns}
\usepackage{bbm}
\usepackage{caption}
\usepackage{subcaption}
\usepackage{xcolor, soul}
\sethlcolor{yellow}

\usepackage[whole]{bxcjkjatype}

\begin{document}
\title{MFBE: Leveraging Multi-Field Information of FAQs for Efficient Dense Retrieval}
\author{
    \textbf{Debopriyo Banerjee$^\dagger$,
    Mausam Jain$^\dagger$,
    Ashish Kulkarni}
    \institute{
        Rakuten Institute of Technology, Rakuten India Enterprise Pvt. Ltd. \\
        \email{\{debopriyo.banerjee, mausam.jain, ashish.kulkarni\}@rakuten.com}
    }
}
\def\thefootnote{$\dagger$}\footnotetext{These authors contributed equally to this work}

\titlerunning{MFBE}
\authorrunning{Banerjee, Jain, Kulkarni}

%
\maketitle
\begin{abstract}
In the domain of question-answering in NLP, the retrieval of Frequently Asked Questions (FAQ) is an important sub-area which is well researched and has been worked upon for many languages. Here, in response to a user query, a retrieval system typically returns the relevant FAQs from a knowledge-base. The efficacy of such a system depends on its ability to establish semantic match between the query and the FAQs in real-time. The task becomes challenging due to the inherent lexical gap between queries and FAQs, lack of sufficient context in FAQ titles, scarcity of labeled data and high retrieval latency. In this work, we propose a bi-encoder-based query-FAQ matching model that leverages multiple combinations of FAQ fields (like, question, answer, and category) both during model training and inference. Our proposed Multi-Field Bi-Encoder (MFBE) model benefits from the additional context resulting from multiple FAQ fields and performs well even with minimal labeled data. We empirically support this claim through experiments on proprietary as well as open-source public datasets in both unsupervised and supervised settings. Our model achieves around 27\% and 23\% better top-1 accuracy for the FAQ retrieval task on internal and open datasets, respectively over the best performing baseline.


\end{abstract}
\section{Introduction}
Customer support (CS) is critical to any business and plays an important role in customer retention, new customer acquisition, branding, and in driving a better experience. In a typical online customer support setting, customers reach out with their queries and are attended to by human agents. This requires businesses to hire and maintain a team of CS agents that scales as a function of the query volume and the productivity of agents that, in turn, translates to operational cost for the business. Customer support automation \cite{mesquita-etal-2022-dense} can help save on this operational cost by providing automated responses to queries and by improving support agent productivity. One of the ways businesses typically try to achieve this is by automatically responding to customer queries from a repository of frequently asked questions (FAQs), thereby, insulating human agents from high query volumes. The success of such a system, measured as the fraction of customer queries that it automatically responds to, then depends on the effectiveness of the user query to FAQ matching.

Given a collection of FAQs where, each FAQ is a multi-field tuple $\langle Q, A, C\rangle$ of question $Q$, answer $A$, and question category $C$, the problem of FAQ retrieval \cite{sakata2019faq,dutta2021sequence,math10081335,karpukhin-etal-2020-dense} is to retrieve the top-$k$ FAQs in response to a user query $q$. Similar to a typical retrieval problem, FAQ retrieval too suffers from the problem of lexical gap between a user query expressed in natural language and the corresponding matching FAQs. This is typically addressed by learning a relevance function between user queries and FAQs using labeled query-FAQ pairs for supervision. Unfortunately, such labeled data is often unavailable or scarce, especially in low-resource settings like Japanese query-FAQ retrieval, which is the domain of our interest. Curating large amounts of such labeled data through manual labeling is often expensive and requires domain knowledge for labeling. 

The main contributions of this paper are summarized as follows:
\begin{itemize}
    \item We propose a bi-encoder based retrieval model~-~MFBE that leverages multi-field information (question, answer and/or categories) in FAQs.
    \item We use different combinations of user query and FAQ fields to create an extended set of pseudo-positive pairs for training. 
    \item We employ multiple FAQ representations during inference for query-FAQ scoring.
\end{itemize}

\section{Related Work}
Question answering~\cite{liu-etal-2021-dense-hierarchical,khattab-etal-2021-relevance} task has been the area of interest in NLP community for a long time and shares the concepts of Information Retrieval (IR) where relevant information from a corpus of documents is retrieved in response to a search query. FAQ retrieval is an example of IR which is the focus of this work. Traditional retrieval methods \cite{DBLP:journals/ftir/RobertsonZ09,Kuzi2020LeveragingSA} mainly depend on lexical features for the retrieval task which limits them to capture the semantics of the query.
In order to address the challenge of lexical gap between user queries and answers, there is a body of work~\cite{bian2021benchmarking,DBLP:journals/corr/abs-2109-02311,DBLP:journals/corr/abs-2106-00882} that trains a semantic retrieval model from labeled data in the form of user queries and matching responses. In recent years, there has been an increasing research~\cite{gao-etal-2021-simcse,liu-etal-2021-fast} on unsupervised learning techniques for text-encoder training eliminating the need for annotated data. 
They 
propose augmentation techniques based on paraphrases of input sentences to generate positive and negative samples for an anchor that are then used to train the retrieval model using a contrastive learning strategy~\cite{chen2020simple}.

The performance of FAQ retrieval task depends upon the (i) choice and design of model architecture and (ii) retrieval and re-ranking algorithms used. 
A combination of bi-encoder and cross-encoder is seen in \cite{liu2021trans,rocketqa_v1}, where the authors start with an unsupervised setting with zero labelled data. Then they iterate between bi-encoder and cross-encoder models generating more labelled samples in each iteration.
This gives a powerful text-encoder model along with annotated dataset. The retriever and the re-ranker can also be jointly trained with the goal of achieving mutual improvement \cite{rocketqa_v2}.


Previous works \cite{sakata2019faq,math10081335,dutta2021sequence} on FAQ retrieval problem focused on 
query-question ($q$-$Q$) similarity using BM25 \cite{DBLP:journals/ftir/RobertsonZ09} and query-answer ($q$-$A$) similarity using BERT \cite{devlin:2019:bert}, where the BERT model parameters are fine-tuned on FAQ question-answer ($Q$-$A$) pairs. 
Sakata {\em et al.} \cite{sakata2019faq} 
(close to our work) employ a two-stage method where they first retrieve a set of FAQs based on $q$-$Q$ similarity and then re-rank these based on $q$-$A$ similarity to obtain the final list of top-$k$ FAQs as response.
Here, question ($Q$) and answer ($A$) are typically referred to as fields. 
Fields can vary depending upon the dataset. For example, Wikipedia page title, content, abstract, etc. have been considered as fields in \cite{liu-etal-2021-dense-hierarchical}.

Dutta {\em et al.} \cite{dutta2021sequence} propose a seq-2seq model for extracting keywords in user queries to identify the intent of a user query for better retrieval of relevant FAQs. Another close work by Assem {\em et al.}~\cite{assem2021dtafa} uses two separate deep learning architectures. They first learn latent lexical relationships between FAQ questions and their paraphrases to generate top-$k$ most relevant similar questions from the collection. These top-$k$ candidates are then fed to an LSTM-based architecture that captures fine-grained differences in semantic context between FAQ questions and their paraphrases thereby improving the accuracy@1. Liu {\em et al.}~\cite{9660603} discuss the difficulty in determining the relevance of query-answer pairs due to their heterogeneity in terms of syntax and semantics and propose to use synthetic data for increasing the positive training examples. Tseng {\em et al.}~\cite{tseng2021faq} cites that existing methods fail to attend to the global information specifically about an FAQ task and propose a graph convolution network-based method to cater to all relations of question and words to generate richer embeddings. They also explore domain specific knowledge graphs for improving question and query representations. 

Some unsupervised sentence embedding methods that closely aligns with our work are \cite{liu-etal-2021-fast} and \cite{gao-etal-2021-simcse}, which are based on contrastive learning using augmentation techniques. Alternate representations of input sentences provide strong positives to the model. \cite{kim2021self} is a self-supervised fine-tuning of BERT, which redesigned the contrastive learning objective to account for different views of the input sentence.

\begin{figure}[t]
\centering
\includegraphics[width=0.7\textwidth]{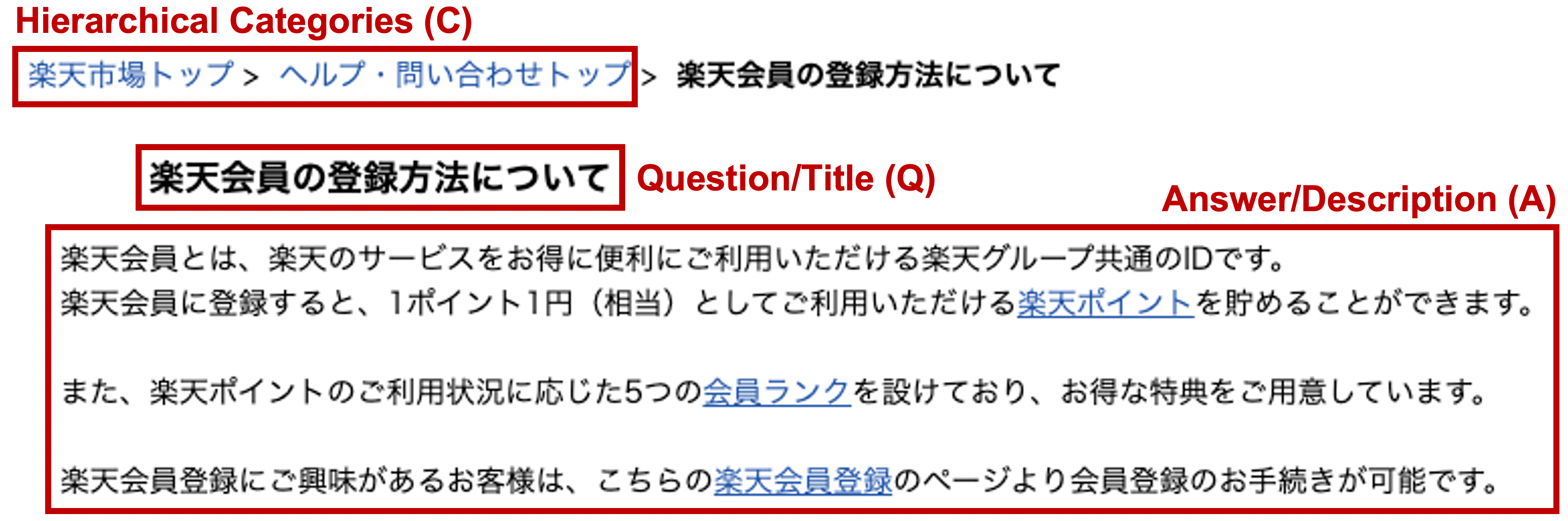}
\caption{Example of an FAQ in Japanese Language (JA).}
\label{fig:faq_example}
\end{figure}

\section{FAQ Retrieval for User Queries}

FAQs are a pre-defined list of $question$-$answer$ pairs available on web portals ({\em e.g.}, banking, e-commerce, telecom, {\em etc.}) that help in addressing user queries without human intervention. In some cases, FAQs are also associated with some hierarchical categories or tags. Fig. \ref{fig:faq_example} shows a sample FAQ that consists of question (or title), answer (or description) and hierarchical categories.
In this paper, we mainly focus on improving the retrieval of FAQs conditioned on user queries using a neural text encoder \cite{Feng:2022:LaBSE}. 
In general, neural text encoders can be categorized into two types: bi-encoders \cite{mazare2018training} 
and cross-encoders \cite{wolf2019transfertransfo}. 
\noindent
\textbf{Bi-Encoder (BE)}: It consists of two encoder branches (with optional weight sharing), where two sentences $S_a$ and $S_b$ are independently passed through each branch, resulting into two sentence embeddings $f_a$ and $f_b$ respectively. 
Their similarity can then be computed using a distance metric like cosine or dot-product of $f_a$ and $f_b$.

\noindent
\textbf{Cross-Encoder (CE)}: The two sentences are first concatenated and then passed through an encoder. The resulting embedding vector is input to a classification head that is typically implemented as a shallow feed forward network. Here, computation of similarity between the two input sentences is modelled as a binary classification task with 1 (0) indicating as similar (not similar). 

Generally, cross-encoders outperform bi-encoders in performance by leveraging the mutual attention among all the words in the concatenated sentence, but they also suffer from high inference latency. The bi-encoder architecture is inherently suited for the FAQ retrieval task as it allows for pre-computation and indexing of sentence embeddings of the FAQs before-hand, which is not possible with cross-encoders. We introduce Multi-Field Bi-Encoders (MFBE) with the aim of improving the performance of bi-encoders for the FAQ retrieval task by leveraging additional context from FAQ titles, description and categories. In this section, we first explain the notations, followed by our proposed approach.

\noindent
\\ \textbf{Notations~-~}Let $\mathcal{F} = \{ F_i\}^N_{i=1}$ be the set of FAQs, where each FAQ $F_i$ consists of a 3-tuple $\langle Q, A, C \rangle$ of question (or title) $Q$, answer (or description) $A$ and categories (or tags) $C$. We define $M$ = \{$Q$, $A$, $QA$, $QC$, $CA$, $QCA$\} as the set of fields, where, each field $m\in M$ is obtained by concatenating one or more of $Q, A$ or/and $C$. Let $M_s = \{Q, QC\} \subset M$ be a subset of fields present in $M$ and $\mathcal{Q}$ be the set of all user queries. 
For a query $q\in \mathcal{Q}$, let $F$ and $\tilde{F}$ denote the matching and non-matching FAQs, respectively, such that ($q$, $F$) forms a matching query FAQ pair and ($q$, $\tilde{F}$) corresponds to a non-matching pair. We denote the field $m$ (or $m_s$) of an FAQ as $F^m$ 
 (or $F^{m_s}$). For example, the field $QA$ of an FAQ is denoted by $F^{QA}$. In this case, for a query $q$, ($q$, $F^{QA}$) and ($q$,$\tilde{F}^{QA}$) are the matching and non-matching query FAQ pairs. We denote the text encoder as $E(\cdot)$, which maps any text to a $d$-dimensional real-valued vector. For every FAQ field $F^m_i$, we compute $E(F^m_i) = f^m_i$ and stack them in a matrix $T_m$, where row $i$ in $T_m$ corresponds to vector $f^m_i$. 
For a query $q$ the corresponding embedding is computed as $E(q)=f_q$. 



\noindent
\\ \textbf{Proposed Approach~-~}We propose to learn the relevance function $rel(x, y; \theta)$ (where $x\in \{q, F^{m_s}\}$; 
when $x=q$, then $y\in \cup_{m\in M} F_m$, else $y \in \cup_{m\in M\setminus M_s} F^m$)
using a pre-trained neural language model \cite{Feng:2022:LaBSE} as the text encoder with $\theta$ as the model parameters. 

\begin{figure}[t]
\centering
\includegraphics[width=\textwidth]{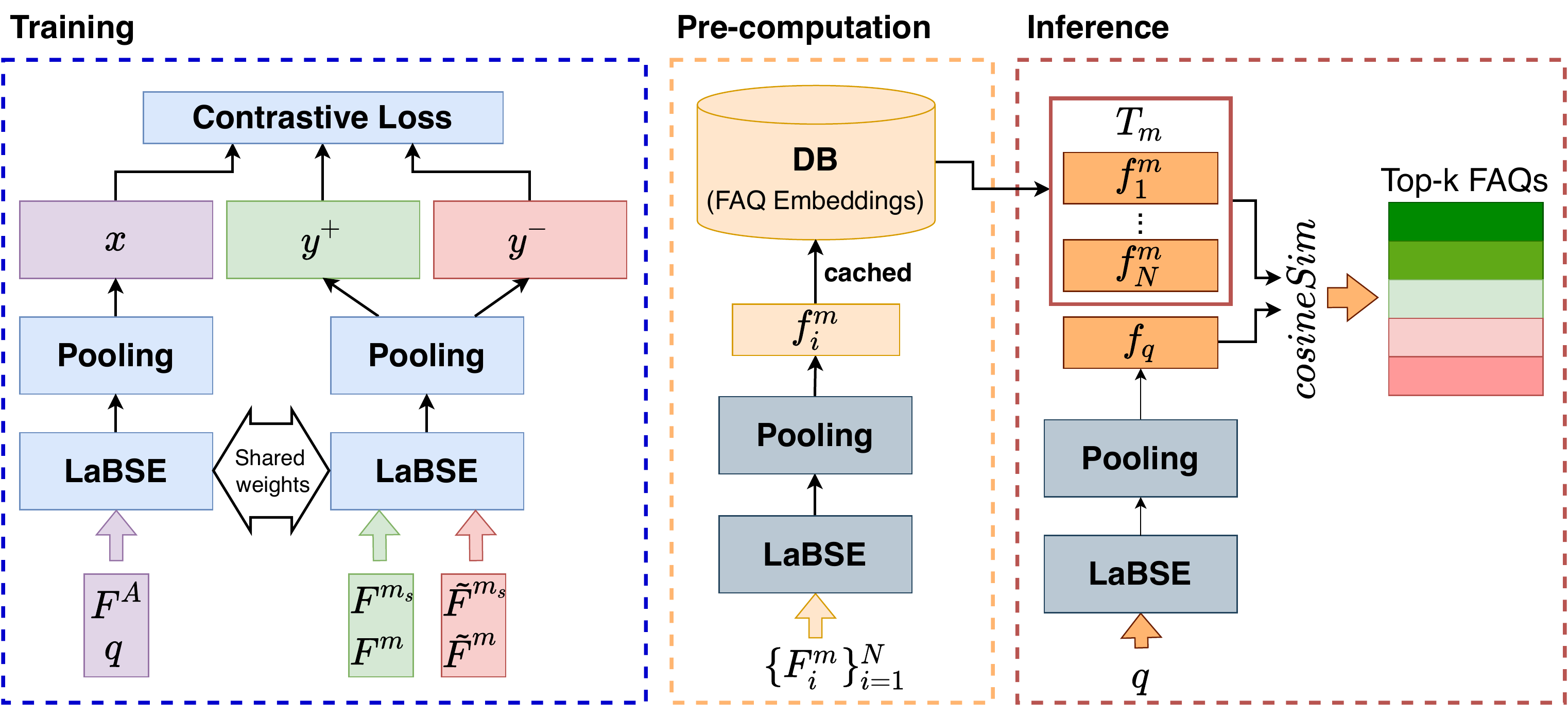}
\caption{\centering Illustration of the working of MFBE model across different stages ~-~ training, pre-computation and inference.}\label{fig:MFBE_overview}
\end{figure}

As shown in Fig. \ref{fig:MFBE_overview}, MFBE consists of Language-agnostic BERT Sentence Embedding (LaBSE) \cite{Feng:2022:LaBSE} model as text encoder in two branches with shared weights.

We compute the similarity between a query $q$ and an FAQ field $F^m$ using the cosine-similarity, i.e., $sim(q, F^m) = cosineSim(f_q, f^m)$.
As mentioned in \cite{karpukhin-etal-2020-dense}, the similarity function should be decomposable and facilitate the pre-computation of representations of the FAQs. L2, inner product and cosine-similarity are some of the widely used similarity functions that are decomposable in nature. We choose cosine-similarity function, which is equivalent to inner product for normalized vectors.

\noindent
\\ \textbf{Training Stage~-~}The goal is to learn a latent space following metric learning \cite{wohlwend-etal-2019-metric}, where matching query and FAQ pairs shall have smaller distance compared to the non-matching pairs. 
Let $\mathcal{D}_{train} = \{(x_i, y^+_i, y^-_{i,1},\ldots,y^-_{i,n})\}^N_{i=1}$ be the training data that consists of $N$ instances, where each instance contains one query or an FAQ field $x_i\in \{q_i, F^{m_s}_i\}$, a matching (positive) FAQ field $y^+_i\in \cup_{m\in M} F^m_i$, when $x_i=q_i$ (or $y^+_i\in \cup_{m\in M\setminus M_s} F^m_i$, when $x_i=F^{m_s}_i$), along with $n$ non-matching (negative) FAQ fields $y^-_{i,j}$. 

We employ the contrastive loss 
function (Eq. \ref{eqn:contrastive_loss}) for fine-tuning the parameters of the MFBE.
\begin{equation}\label{eqn:contrastive_loss}
    L(x_i, y^+_i,y^-_{i,1},\ldots, y^-_{i,n}) = -log\frac{e^{sim(x_i, y^+_i)}}{e^{sim(x_i, y^+_i)} + \sum_j e^{sim(x_i, y^-_{i,j})}}
\end{equation}

We train three variants of unsupervised and supervised MFBE models~-~MFBE$_{unsup}$, MFBE$_{sup}$, and MFBE$_{sup^*}$.
Triplets used for each variant are as follows:

\begin{equation}
    triplets_{unsup} = \bigcup_{m_s\in M_s}\{(F^A, F^{m_s}, \tilde{F}^{m_s}), (F^{m_s}, F^A, \tilde{F}^{m_s})\}
\end{equation}

\begin{equation}
    triplets_{sup} = \bigcup_{m\in M} \{(q, F^m, \tilde{F}^m), (F^m, q, \tilde{F}^m)\}
\end{equation}

\begin{equation}
    triplets_{sup^*} = triplets_{unsup} \cup triplet_{sup}
\end{equation}

\noindent
\textbf{Positive and Negative FAQs}~-~In FAQ datasets, positive examples are explicitly present from the manual annotation of query-FAQ, i.e., ($q, F^m$) pairs. For boosting the number of training samples, we consider both the pairs ($F^Q, F^A$) and ($F^{QC}, F^A$) from an FAQ as proxy for $(q, F^m)$. But, negative examples are not explicitly present. However, the choice of negative samples play a decisive role in learning an effective text encoder. So, we consider \textit{Gold} negatives  \cite{karpukhin-etal-2020-dense}, i.e., positive FAQs paired with other non-matching queries that appear in the training set.
In order to make the training computation more efficient, we make use of \textit{Gold} FAQs from the same mini-batch as negatives, termed as in-batch negatives \cite{karpukhin-etal-2020-dense}. In addition, we consider one negative FAQ sample for each matching query-FAQ pair. 

\noindent\textbf{Pre-computation of FAQ Embeddings~-~}After training the text-encoder $E(\cdot)$, we use it to pre-compute and store $T_{m}$ (Eq. \ref{eqn:faq_embeddings}) corresponding to the field $m$. We use the best performing field $m^{best} \in M$ (Eq. \ref{eqn:best_field}), in terms of $Acc$ (accuracy@1, Eq. \ref{eqn:A@1}), evaluated on the test set $\mathcal{D}_{test}$ for the query-FAQ matching task. 
\begin{equation}\label{eqn:faq_embeddings}
T_{m} = \begin{bmatrix}
            f^{m}_1 f^{m}_2 \hdots f^{m}_N
        \end{bmatrix}^T
\end{equation}
where $f^{m}_i = E(F^{m}_i)$, for $i=1,2,\ldots, N$.
\begin{equation}\label{eqn:best_field}
    m^{best} = \underset{m\in M}{\arg\max} 
 \text{ }Acc(\mathcal{D}_{test})
\end{equation}
In Eq. \ref{eqn:best_field}, test set $\mathcal{D}_{test} = \{ (q, F) | F\in \mathcal{F} \text{ is a matching FAQ for query } q\in \mathcal{Q}_{test}\}$, where $\mathcal{Q}_{test} \subset \mathcal{Q}$. As each FAQ consists of different FAQ fields $F^m$ ($\forall m \in M$), we define $Acc(\cdot)$ as below.
\begin{equation}\label{eqn:A@1}
    Acc(\mathcal{D}_{test}) = \frac{1}{|\mathcal{Q}_{test}|}\sum_{q\in \mathcal{Q}_{test}}\mathbbm{1}[\underset{i}{\{\arg\max}\text{ }cosineSim(f_q, f^m_i)\} \cap gt(q) \neq \Phi]
\end{equation}
Here, $gt(q)$ denotes the set of FAQ indexes present in the ground truth labels corresponding to query $q$. 

\noindent
\textbf{Inference~-~} During inference stage, for a given query $q$, we first compute the input query embedding $f_q$. Then, we calculate cosine similarity scores of $f_q$ and row vectors of $T_{m^{best}}$ (which are pre-computed embeddings of FAQs) and return the top-k candidates, sorted in descending order of their scores.

\section{Experiments, Implementation and Results}

\bgroup
\def\arraystretch{1}%
\begin{table}[t]
    \scriptsize
    \caption{\footnotesize \centering Datasets used in this paper. For open datasets queries are separated into five folds. (JA=Japanese, EN=English)}
    \centering
    \begin{tabular}{c c c c c c c c c c c c c c}
        \toprule
        \multirow{2}{*}{\bf{Type}} & & 
        \multirow{2}{*}{\bf{Name}} & & 
        \multirow{2}{*}{\bf{Language}} & & 
        \multirow{2}{*}{\bf{\#FAQs}} & & 
        \multicolumn{2}{c}{\bf{\#Queries}} & &
        \multicolumn{3}{c}{\bf{\#Avg. sentence length}} \\
        \cmidrule{9-10} 
        \cmidrule{12-14} 
        & & & & & & & & Train & Test & & Query & FAQ-Q & FAQ-A \\
        \cmidrule{1-14} 
        \multirow{4}{*}{Internal} 
        & & IDS1 & & JA & & 795 & & 1782 & 825 & & 28.5 & 24.8 & 166.6 \\
        & & IDS2 & & JA & & 510 & & 481 & 139 & & 28.6 & 31.3 & 514.6 \\
        & & IDS3 & & JA & & 1129 & & 528 & 152 & & 14.9 & 33.1 & 473.8 \\
        & & IDS4 & & JA & & 661 & & 505 & 145 & & 17.6 & 33.4 & 471.8 \\
        \cmidrule{1-14}
        \multirow{3}{*}{Open} 
        & & LocalGov & & JA & & 1786 & & \multicolumn{2}{c}{749} & & 26.1 & 31.1 & 357.2 \\
        & & Stack-FAQ & & EN & & 125 & & \multicolumn{2}{c}{1249} & & 73.3 & 55.7 & 513.8 \\
        & & COUGH & & EN & & 7115 & & \multicolumn{2}{c}{1201} & & 74.4 & 76.5 & 711.7 \\
        \bottomrule
    \end{tabular}
    \label{table:datasets_used}
\end{table}

\noindent
\textbf{Datasets~-~}We conduct experiments on both internal proprietary and open datasets. Table \ref{table:datasets_used} shows the list of datasets and their details used in this work.

\noindent
\textbf{Internal datasets~-~}IDS1, IDS2, IDS3 and IDS4 are Japanese language company internal datasets related to e-commerce, leisure, communication and payment domains respectively. These are carefully prepared by majority voting by five native Japanese speakers across query-FAQ annotations.

\noindent
\textbf{Open datasets~-~}
\textit{LocalGov}, introduced in \cite{sakata2019faq}, is a Japanese language dataset which is constructed from Japanese administrative municipality domain. \textit{Stack-FAQ} is described in \cite{KARAN2018418} as an English language dataset prepared from threads in StackExchange website concerning web apps domain. \textit{COUGH} is another English dataset \cite{zhang2021cough} constructed by scraping data from 55 websites (like, CDC and WHO) containing user queries and FAQs about Covid-19.


\noindent
\\ \textbf{Implementation Details~-~}We use single NVIDIA A100 GPU with 40G VRAM for all experiments. In all our experiments, we set maximum sequence length as 256 (for training and testing) and batch size as 32 which are constrained by the choice of GPU. For all datasets, the number of training epochs is set as 15. The choice of optimizer, learning rate and embedding dimension follows from \cite{reimers-gurevych-2019-sentence} for all the experiments except for multi-domain fine-tuning experiments (Table \ref{table:cross_multi_domain_res} last 4 rows) where learning rate is set as $2e-7$. We set weight decay $\omega = 1e-5$ by employing grid search between 1e-1 and 1e-10 (reducing by 0.1x). For each matching query-FAQ pair, we consider 10 negative FAQs (i.e., n=10 after experimenting with other values, such as 5, 10, and 20).
All our baseline experiments follow same settings as the proposed models and are initialized with LaBSE checkpoint\footnote{https://huggingface.co/sentence-transformers/LaBSE/tree/main}. Code and relevant material of this work can be found here\footnote{https://github.com/mausamsion/MFBE}.

\begin{table}[t]
    \scriptsize
    \centering
    \caption{\footnotesize \centering Results on internal datasets.}
    \begin{tabular}{l c c c c c c c c c c c c c c c c}
        \toprule
        \multirow{2}{*}{\bf{Model}} & &
            \multicolumn{3}{c}{\bf{IDS1}} & & \multicolumn{3}{c}{\bf{IDS2}} & & \multicolumn{3}{c}{\bf{IDS3}} & & \multicolumn{3}{c}{\bf{IDS4}}\\
            \cmidrule{3-5} 
            \cmidrule{7-9} 
            \cmidrule{11-13}
            \cmidrule{15-17}
        & & Acc & MRR & NDCG & & Acc & MRR & NDCG & & Acc & MRR & NDCG & & Acc & MRR & NDCG \\
        \cmidrule{1-17}
        \textit{Baselines} & & & & & & & & & & & & & \\
        BM25 \cite{DBLP:journals/ftir/RobertsonZ09} & & 23.5 & 29.8 & 30.6 & & 36.0 & 43.2 & 30.3 & & 38.2 & 47.3 & 31.7 & & 35.9 & 46.0 & 31.8 \\
        LaBSE \cite{Feng:2022:LaBSE} & & 29.9 & 38.6 & 40.1 & & 36.0 & 48.4 & 35.8 & & 49.3 & 59.2 & 41.7 & & 51.0 & 59.0 & 40.4 \\
        DPR \cite{karpukhin-etal-2020-dense} & & 41.5 & 51.7 & 20.5 & & 47.5 & 55.9 & 34.6 & & 46.7 & 57.9 & 37.0 & & 46.9 & 57.9 & 35.3 \\
        \cmidrule{1-17}
        \textit{Proposed} & & & & & & & & & & & & & \\
        MFBE$_{unsup}$ & & 33.2 & 43.9 & 46.2 & & 51.1 & 60.3 & 44.7 & & 51.3 & 59.7 & 43.8 & & 51.0& 61.7& 43.3\\
        MFBE$_{sup}$ & & 51.0 & 60.6 & 62.4 &  & 60.4 & \bf{71.3} & \bf{58.1} & & 57.9 & 67.8 & 52.8 & & 64.1 & 72.2 & 54.9 \\
        MFBE$_{{sup}^*}$ & & \bf{59.2} & \bf{66.4} & \bf{66.8} &  & \bf{61.2} & 69.8 & 55.1 & & \bf{57.9} & \bf{68.9} & \bf{53.9} & & \bf{65.5} & \bf{74.9} & \bf{57.8}\\
        \bottomrule
    \end{tabular}
    \label{table:internal_ds_res}
\end{table}

\begin{table}[t]
    \scriptsize
    \centering
    \caption{\footnotesize \centering Results on open datasets.}
    \begin{tabular}{l c c c c c c c c c c c c}
        \toprule
        \multirow{2}{*}{\bf{Model}} & &
            \multicolumn{3}{c}{\bf{LocalGov}} & & \multicolumn{3}{c}{\bf{Stack-FAQ}} & & \multicolumn{3}{c}{\bf{COUGH}} \\
            \cmidrule{3-5} 
            \cmidrule{7-9} 
            \cmidrule{11-13}
        & & Acc & MRR & NDCG & & Acc & MRR & NDCG & & Acc & MRR & NDCG \\
        \cmidrule{1-13}
        \textit{Baselines} & & & & & & & & & & & & \\
        BM25 \cite{DBLP:journals/ftir/RobertsonZ09} & & 26.4 & 33.4 & 26.5 & & 40.0 & 49.1 & 52.9 & & 39.1 & 48.0 & 26.1\\
        LaBSE \cite{Feng:2022:LaBSE} & & 26.3 & 35.9 & 28.8 & & 43.6 & 55.6 & 60.6 & & 23.0 & 31.7 & 17.7 \\
        DPR \cite{karpukhin-etal-2020-dense} & & 46.5 & 55.2 & 44.7 & & 88.8 & 91.9 & 93.0 & & 42.8 & 51.5 & 29.3 \\
        \cmidrule{1-13}
        \textit{Proposed} & & & & & & & & & & & & \\
        MFBE$_{unsup}$ & & 53.8 & 63.5 & 54.9 & & 85.9 & 91.0 & 92.8 & & 46.0 & 57.7 & 35.5 \\
        MFBE$_{sup}$ & & 61.3 &  72.0 & 64.8 & &  \textbf{98.0} &  \textbf{98.8} &  \textbf{99.0} & & \textbf{53.1}  &  \textbf{65.2} &  \textbf{40.3} \\
        MFBE$_{{sup}^*}$ & & \textbf{62.9} & \textbf{72.3} & \textbf{65.7} & & 96.5 & 98.0 & 98.5 & & 50.1 & 61.0 & 37.7 \\
        \bottomrule
    \end{tabular}
    \label{table:open_ds_res}
\end{table}

\begin{table}[t]
    \scriptsize
    \centering
    \caption{\footnotesize \centering Results of cross- and multi-domain experiments using MFBE$_{sup^*}$ model.}    
    \begin{tabular}{l c c c c c c c c c c c c c c c c}
        \toprule
        \multirow{2}{*}{\bf{Model}} & &
            \multicolumn{3}{c}{\bf{IDS1}} & & \multicolumn{3}{c}{\bf{IDS2}} & & \multicolumn{3}{c}{\bf{IDS3}} & & \multicolumn{3}{c}{\bf{IDS4}}\\
            \cmidrule{3-5} 
            \cmidrule{7-9} 
            \cmidrule{11-13}
            \cmidrule{15-17}
        & & Acc & MRR & NDCG & & Acc & MRR & NDCG & & Acc & MRR & NDCG & & Acc & MRR & NDCG \\
        \cmidrule{1-17}
        \multicolumn{17}{l}{\textit{Cross-domain}} \\
        IDS$_1$ & & - & - & - & & 47.5 & 62.2 & 55.2 & & 55.3  & 66.0 & 53.0 & & 53.1 & 65.7 & 53.6 \\
        IDS$_2$ & & 37.3 & 48.3 & 50.3 & & - & - & - & & 47.4 & 56.9 & 39.2 & & 48.3 & 58.3 & 40.5 \\
        IDS$_3$ & & 42.4 & 53.1 & 55.5 & & 44.6 & 57.0 & 44.2 & & - & - & - & & 57.2 & 63.6 & 45.8 \\
        IDS$_4$ & & 40.5 & 51.9 & 54.3 & & 44.6 & 58.5 & 44.8 & & 50.0 & 61.4 & 46.1 & & - & - & - \\
        \cmidrule{1-17}
        \multicolumn{17}{l}{\textit{Multi-domain}} \\
        IDS$^*$ & & 55.5 & 64.2 & 65.5 & & 62.6 & 71.2 & 54.8 & & 55.3 & 66.2 & 50.8 & & 59.3 & 67.9 & 50.8 \\
        \cmidrule{1-17}
        \multicolumn{17}{l}{\textit{Cross and Multi-domain}} \\
        IDS$^*_1$ & & 55.0 & 64.0 & 65.2& & 62.6 & 71.3 & 54.9 & & 55.3 & 66.2 & 50.7 & & 59.3 & 68.0 & 50.7 \\
        IDS$^*_2$ & & 54.4 & 63.3 & 64.7 & & 63.3 &	71.9 & 55.4 &  & 54.6 & 65.6 & 50.6 & & 59.3 & 68.2 & 51.6 \\
        IDS$^*_3$ & & 54.8 & 63.8 & 65.1 & & 61.9 & 70.5 & 54.5 & & 55.9 & 67.1	& 51.4& & 57.9 & 67.2 & 51.0 \\
        IDS$^*_4$ & & 54.2 & 63.0 & 64.2 & & 61.2 & 70.3 & 53.4 & & 56.6 & 67.0 & 50.6 & & 61.4 & 69.8	& 52.9 \\
        \bottomrule
    \end{tabular}
    \label{table:cross_multi_domain_res}
\end{table}


\noindent
\\ \textbf{Results and Discussion~-~}We report accuracy@1 (Acc), mean reciprocal rank @5 (MRR) and normalized discounted cumulative gain @5 (NDCG) and compare our proposed models with multiple baselines, such as, BM25 (lexical feature-based IR model), LaBSE (heavy-weight dense multilingual text encoder), and DPR (dense bi-encoder with independently learned encoders). 
Across all baselines, we keep the train settings related to field combinations same as our MFBE$_{sup^*}$ model.
From Table \ref{table:internal_ds_res}, BM25 is the worst performing model, because of dependence on lexical features, hence fails to capture semantics among queries and FAQs. MFBE$_{sup^*}$ outperforms all baselines across different datasets.
In Table \ref{table:open_ds_res}, we report 5-fold cross validation results. MFBE$_{sup}$ and MFBE$_{sup^*}$ shows the best performance across all the baselines. MFBE$_{unsup}$, which has zero query knowledge, performs better than DPR, which is the best performing baseline of all, in two datasets out of three. Table \ref{table:example_input_output} illustrates sample input query and output top-1 prediction of DPR and MFBE$_{unsup}$ models where we see that DPR fails to capture the semantic meaning of the input thus returning irrelevant response.

\noindent
\textbf{Cross-domain~-~}Table \ref{table:cross_multi_domain_res} shows the cross-domain results of MFBE$_{sup^*}$ where the model is trained on one dataset and evaluated on completely unknown ones (zero-shot setting). For example, the first row corresponds to the case where MFBE$_{sup^*}$ is trained on IDS1 and evaluated on IDS2, IDS3, and IDS4.
We observe that the zero-shot performance of MFBE$_{sup^*}$ is the best, when trained on IDS1, compared to others. This is because IDS1 consists of a large number of labelled user queries (nearly 1.7k) in the train split. 
\\
\textbf{Multi-domain~-~}Here MFBE$_{sup^*}$ model is trained on all the internal datasets IDS[1-4] denoted as IDS$^*$. The average drop in Acc is only 4\%, compared to the last row of Table \ref{table:internal_ds_res}. The model trained on IDS$^*$ is more robust across multiple domains with better performance in some cases (e.g., IDS2), making it useful for leveraging cross-domain knowledge. \\
\textbf{Cross and Multi-domain~-~}In this case, the model is first trained on IDS$^*$, then fine-tuned on one dataset, and finally evaluated on all datasets. For example, the sixth row corresponds to the case, where MFBE$_{sup^*}$ is first trained on IDS$^*$, then fine-tuned on IDS1 (denoted as IDS$^*_1$), and finally evaluated on all the datasets, i.e., IDS[1-4] (same for the last three rows). There is no significant change in performance when compared to results of fifth row because of prior exposure to the corresponding datasets.

\begin{table}[t]
    \scriptsize
    \caption{\footnotesize \centering Example input and outputs from COUGH (EN) and LocalGov (JA) datasets of two baseline models and our MFBE$_{sup^*}$ model. In all the examples, MFBE$_{sup^*}$ returns the most relevant response in top-1.}
    \centering
    \def\arraystretch{1.5}
    \setlength\tabcolsep{0.12cm}
    \begin{tabularx}{\textwidth}[t]{X X X} 
        \toprule
        \multirow{2}{*}{\bf{Input}} & \multicolumn{2}{c}{\bf{Top FAQ (question) prediction}} \\
        \cmidrule{2-3}
        & DPR & MFBE$_{sup^*}$ \\
        
        \cmidrule{1-3}
        
        Is personal protective equipment sufficient to protect others? & 
        Are there exemptions to who has to wear a face covering? & 
        Are cloth face coverings the same as personal protective equipment (PPE)? \\
        
        How should i adjust my feeling during pendemic period? & 
        I traveled and have been sick ever since I got back. What should I do? & 
        During this time, it is important to be S.M.A.R.T. about staying active \\
        
        \cmidrule{1-3}
        
        国民年金の納付書を誤って捨ててしまいました。どうしたらいいでしょうか？ & 
        【児童手当現況届】間違って記入した場合は、どうしたらいいですか。 & 
        国民年金保険料を支払いたいのですが納付書をなくしてしまいました。 \\

        納税証明書が必要なのですが、どこで入手できますか？ & 
        介護保険料の納付書を紛失してしまった。再交付してほしいのですが？ & 
        納税証明書（法人市民税、事業所税を除く）を取得したい。 \\
        
        \bottomrule
    \end{tabularx} 
    \label{table:example_input_output}
\end{table}

\begin{figure}[t]
\centering
\includegraphics[width=\textwidth]{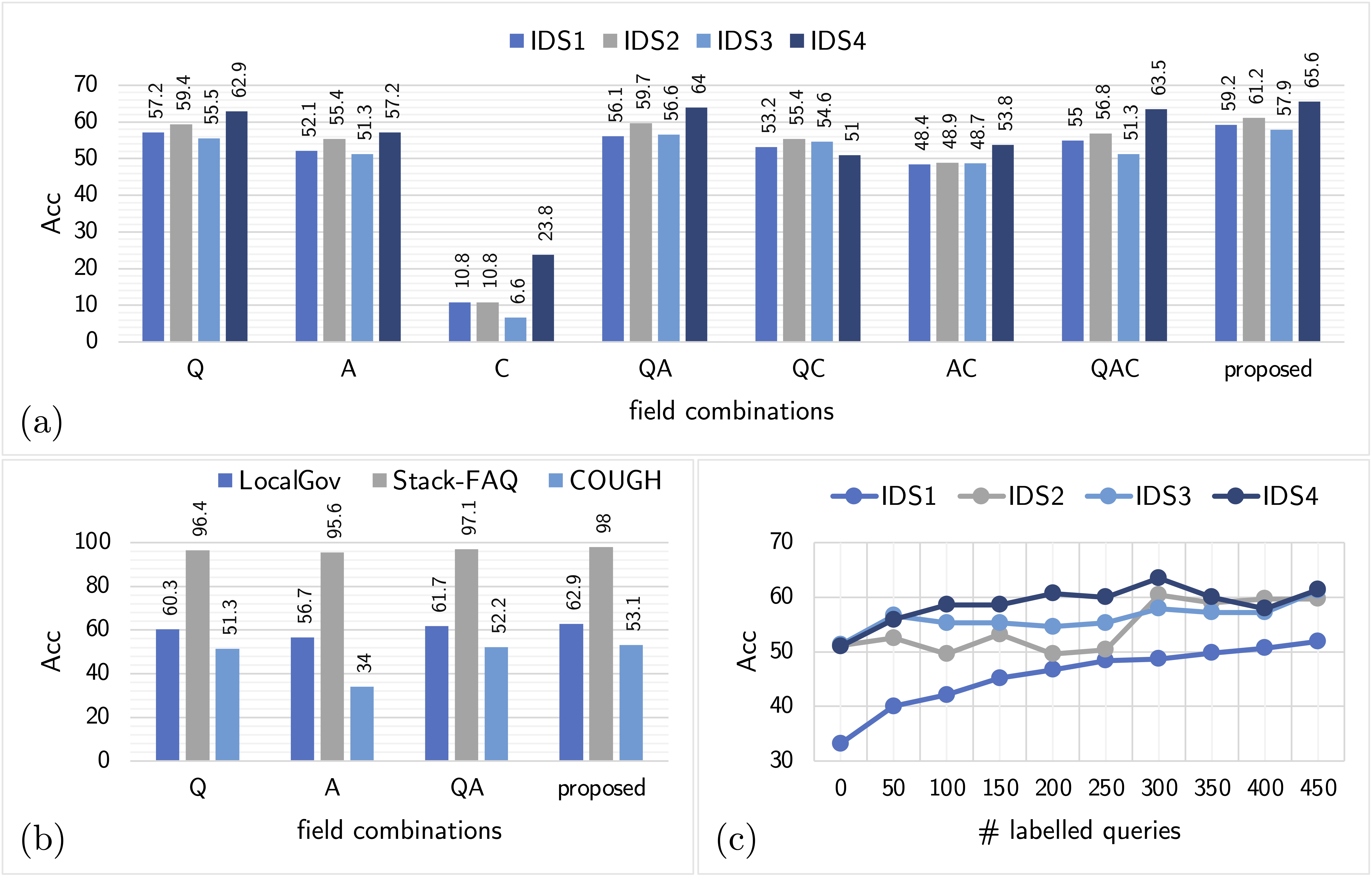}
\caption{\footnotesize \centering Ablation experiments with MFBE$_{sup^*}$ model. (a) Variation across multi-field combinations on internal datasets (b) Variation across multi-field combinations on open datasets (c) Variation in the number of training query-FAQ pairs}
\label{fig:ablation}
\end{figure}

\noindent
\textbf{Ablation experiments~-~}We train MFBE$_{sup^*}$, $\forall m\in M$ (taking one  at a time) and varying the number of labelled queries.
In figure \ref{fig:ablation}-(a) and (b) we show the performance of our model as FAQ field combinations are changed (which consistent at both training and testing). It is observed that using category information adds noise and degrades performance which can be due to the inefficient usage of this field. The category field has keywords and hierarchy which needs to be leveraged but in this work, for simplicity, we concatenated these keywords to other input fields making it as a part of input string. The 'proposed' numbers are the best numbers across all of our proposed models (as discussed in Table \ref{table:internal_ds_res} and \ref{table:open_ds_res}).
From figure \ref{fig:ablation}-(c) it is observed that our MFBE$_{sup^*}$ model is a good candidate for the scenarios where there is less annotated data with the accuracy flattening after around 300 query-FAQ pairs. This makes it suitable for bootstrapping to new domains where there are FAQ documents and no or less query-FAQ pairs.



\section{Conclusion}
In this paper, we proposed MFBE, a bi-encoder based retrieval model that make use of information from multiple fields in FAQs to improve the text embedding quality and thus better sentence matching. We also create an extended set of pseudo-positive training pairs by using various combinations of user-query and FAQ fields. Then we use these multiple FAQ representations to make inference on input queries. Our model outperforms the baselines by 27\% and 23\% (in terms of accuracy@1) on internal and open-datasets, respectively. Cross-domain experiment results for the MFBE$_{sup^*}$ model over our internal datasets shows the potential of this kind of proposed approach to be useful in cold-start settings, which is common in real-world scenarios. Also, multi-domain experiment proves the possibility of multi-domain knowledge sharing using a single model which performs good across most of the datasets it is trained on. We also do ablation on semi-supervised setting (queries variation) and effect of FAQ field combinations. 


%
%
%
\bibliographystyle{splncs04}
\bibliography{ref}
\end{document}